\documentclass[pra,twocolumn,aps,floatfix,showpacs,tightenlines,superscriptaddress,amsmath,amssymb,showkeys,10pt,nofootinbib]{revtex4-2}
\usepackage{slashed}
\usepackage{xcolor}
\usepackage{graphicx}
\usepackage{hyperref}

\newcommand{\be}{\begin{equation}}\newcommand{\ee}{\end{equation}}
\newcommand{\bea}{\begin{eqnarray}}\newcommand{\eea}{\end{eqnarray}}
\newcommand{\brr}{\begin{array}}\newcommand{\err}{\end{array}}
\newcommand{\bit}{\begin{itemize}}\newcommand{\eit}{\end{itemize}}
\newcommand{\ben}{\begin{enumerate}}\newcommand{\een}{\end{enumerate}}
\newcommand{\bib}{\bibitem}
\newcommand{\bbm}{\begin{bmatrix}}\newcommand{\ebm}{\end{bmatrix}}
\newcommand{\ba}{\begin{array}}
\newcommand{\ea}{\end{array}}
\newcommand{\G}{\textbf}
\newcommand{\C}{\textit}
\newtheorem{mydef}{Definition}
\newtheorem{Lemma}{Lemma}
\newcommand{\bd}{\begin{mydef}} \newcommand{\ed}{\end{mydef}}
\newcommand{\bthe}{\begin{theorem}} \newcommand{\ethe}{\end{theorem}}
\newcommand{\ble}{\begin{Lemma}} \newcommand{\ele}{\end{Lemma}}
\newcommand{\ul}{\underline}
\newcommand{\mc}{\mathcal}
\newcommand{\mr}{\mathrm}
\newcommand{\dr}{\mathrm{d}}
\newcommand{\revision}{\cdred}
\newcommand{\ommision}{\cdblue}
\def\ha{\frac{1}{2}}
\def\lag{\mathcal{L}}
\def\intx{\int \!\!\mathrm{d}^3 {\G x}}
\def\intk{\int \!\!\mathrm{d}^3 {\G k}}
\def\tr{\mathrm{tr}}
\def\Tr{\mathrm{Tr}}
\def\ph{\varphi}
\def\mpsi{\boldsymbol{\psi}}
\def\mpsib{\overline{\boldsymbol{\psi}}}
\def\mnu{\boldsymbol{\nu}}
\def\mnub{\overline{\boldsymbol{\nu}}}
\def\Dia{\Diamond}\def\lab{\label}\def\lan{\langle}\def\lar{\leftarrow}
\def\lf{\left}\def\lrar{\leftrightarrow}
\def\Lrar{\Leftrightarrow}\def\noi{\noindent}
\def\non{\nonumber}\def\ot{\otimes}\def\pa{\partial}\def\ran{\rangle}
\def\rar{\rightarrow}\def\Rar{\Rightarrow}
\def\ri{\right}\def\ti{\tilde}\def\we{\wedge}\def\wti{\widetilde}
\def\al{\alpha}\def\bt{\beta}\def\ga{\gamma}\def\Ga{\Gamma}
\def\de{\delta}\def\De{\Delta}\def\ep{\epsilon}
\def\ze{\zeta}\def\te{\theta}\def\ka{\kappa}
\def\la{\lambda}\def\La{\Lambda}\def\si{\sigma}\def\Si{\Sigma}
\def\om{\omega}\def\Om{\Omega}
\def\AB{{_{A,B}}}\newcommand{\mlab}[1]{\label{#1}}
\def\CP{{_{C\!P}}}\def\T{{_{T}}}
\def\AB{{_{A,B}}}\def\mass{{_{1,2}}}
\def\flav{{e,\mu}}\def\1{{_{1}}}\def\2{{_{2}}}
\def\bp{{\bf {p}}}\def\bk{{\bf {k}}}\def\br{{\bf {r}}}\def\bx{{\bf {x}}}
\def\by{{\bf {y}}}\def\bl{{\bf {l}}}\def\bq{{\bf {q}}}\def\bj{{\bf {j}}}
\def \ak{\alpha^r_{{\bf k},e}(0)}\def \akd{\alpha^{r\dag}_{{\bf k},e}(0)}
\def\ap{\alpha^s_{{\bf p},e}(0)}\def\apd{\alpha^{s\dag}_{{\bf p},e}(0)}
\def\br{{\bf{r}}}\def\bI{{\bf{I}}}
\def\fourint{\int\!\!\!\int\!\!\!\int\!\!\!\int}\def\threeint{\int\!\!\!\int\!\!\!\int}
\def\twoint{\int\!\!\!\int}
\def\CKM{\tiny CKM}
\def\Qo{{_{Q\!,1}}}\def\Qt{{_{Q\!,2}}}\def\ko{{_{k\!,1}}}\def\kt{{_{k\!,2}}}
\newcommand{\ide}{1\hspace{-1mm}{\rm I}}
\newcommand{\noH}{:\;\!\!\;\!\!:H:\;\!\!\;\!\!:}
\def\noHe0{:\;\!\!\;\!\!:H_e(0):\;\!\!\;\!\!:}
\def\noHm0{:\;\!\!\;\!\!:H_\mu(0):\;\!\!\;\!\!:}
\def\nof{:\;\!\!\;\!\!:}
\def\vect#1{{\bm #1}}
\def \r{\mathrm}
\def\Dia{\Diamond}\def\lab{\label}
\def\lan{\langle}\def\lar{\leftarrow}
\def\lf{\left}\def\lrar{\leftrightarrow}
\def\Lrar{\Leftrightarrow}\def\noi{\noindent}
\def\non{\nonumber}\def\ot{\otimes}
\def\pa{\partial}\def\ran{\rangle}
\def\rar{\rightarrow}\def\Rar{\Rightarrow}
\def\ri{\right}\def\ti{\tilde}
\def\we{\wedge}\def\wti{\widetilde}
\def\al{\alpha}\def\bt{\beta}\def\ga{\gamma}
\def\Ga{\Gamma}\def\de{\delta}\def\De{\Delta}
\def\ep{\epsilon}\def\ze{\zeta}\def\te{\theta}
\def\ka{\kappa}\def\la{\lambda}
\def\La{\Lambda}\def\si{\sigma}\def\Si{\Sigma}
\def\om{\omega}\def\Om{\Omega}
\def\AB{{_{A,B}}}
\def\CP{{_{C\!P}}}\def\T{{_{T}}}
\def\AB{{_{A,B}}}\def\mass{{_{1,2}}}
\def\flav{{e,\mu}}\def\1{{_{1}}}\def\2{{_{2}}}
\def\nof{:\;\!\!\;\!\!:}
\def\wQ{Q}
\def\wwQ{Q}

\def\I{{_{\rm{I}}}}\def\II{{_{\rm{II}}}}
\def\A{{_{A}}}\def\B{{_{B}}}

\def\rran{\ran\!\ran}
\def\llan{\lan\!\lan}

\begin{document}

\title{Neutrino decoherence and violation of the strong equivalence principle}

\author{Luca Buoninfante}
\email{luca.buoninfante@su.se}
\affiliation{Nordita, KTH Royal Institute of Technology and Stockholm University, Hannes Alfv\'ens v\"ag 12,
Stockholm, SE-106 91, Sweden}
\author{Giuseppe Gaetano Luciano}
\email{giuseppegaetano.luciano@udl.cat}
\affiliation{Applied Physics Section of Environmental Science Department,
Escola Polit\`ecnica Superior, Universitat de Lleida, Av. Jaume II, 69, 25001 Lleida, Spain}
\author{Luciano Petruzziello}
\email{lupetruzziello@unisa.it}
\affiliation{INFN Sezione di Napoli, Gruppo collegato di Salerno, Italy}
\affiliation{Dipartimento di Ingegneria Industriale, Universit\`a di Salerno,
Via Giovanni Paolo II 132, 84084 Fisciano (SA), Italy}
\affiliation{Institut f\"ur Theoretische Physik, Albert-Einstein-Allee 11, Universit\"at Ulm, 89069 Ulm, Germany}
\author{Luca Smaldone}
\email{lsmaldone@unisa.it}
\affiliation{INFN Sezione di Napoli, Gruppo collegato di Salerno, Italy}
\affiliation{Dipartimento di Fisica, Universit\`a di Salerno, Via Giovanni Paolo II, 132 84084 Fisciano, Italy}

\begin{abstract}
We analyze the dynamics of neutrino Gaussian wave-packets, the damping of flavor oscillations and decoherence effects within the framework of extended theories of gravity. We show that, when the underlying description of the gravitational interaction admits a violation of the strong equivalence principle, the parameter quantifying such a violation modulates the wave-packet spreading, giving rise to potentially measurable effects in future neutrino experiments. 
\end{abstract}

\maketitle

\section{Introduction}
 
A full-fledged description of neutrino theory and phenomenology represents one of the main challenges in elementary particle physics today. 
Because of the very elusive nature of neutrinos, experimental tests turn out to be particularly demanding, as also witnessed by the nearly thirty years elapsed between the prediction of neutrino by Pauli and the first-ever confirmation of its existence~\cite{reines,reines2}. 
In spite of these limitations, 
a theoretical treatment of neutrino physics did not take a long time to be formalized~\cite{pontecorvo,pontecorvo2,pontecorvo3,pontecorvo4,giuntibook}. 
In the pioneering works by Pontecorvo~\cite{pontecorvo,pontecorvo2}, it was understood that the physical (i.e. interacting) flavor states of neutrino are not pure mass (i.e. Hamiltonian) eigenstates, but a mixture of the latter.
This gives rise to the phenomena of neutrino mixing and oscillations, which have by now been experimentally tested with accurate precision~\cite{kamio,kamio2,kamio3,kamio4,Exp1,Exp3,Exp3} and provide the first solid evidence of physics beyond the Standard Model.
Further progress towards revealing neutrino
properties has been recently achieved in the context of quantum field theory~\cite{QFT1,QFT2,QFT3,QFT4}.

A very active research in the above framework deals with
the interplay between neutrinos and gravity. Since the early study~\cite{stodolsky}, great effort has been devoted to this subject,  
due to the peculiar features of neutrinos
as a unique source of information for the fine-grained analysis of the gravitational interaction~\cite{cardall,fornengo,fornengo2,ahluwburgard,piriz,kojima,punzi,capoz1,ourepl,lamb1,lamb2,ourextended,role,matsas,capolupo,
neutrinounruh,capolupo2,cpvio,frontiers,Lucianonon,dvornikov,capolupo3,dvornikov2,timeenergy,swami,capolupo4,dvornikov4} (for a recent review, see~\cite{ourreview}). For instance, in~\cite{fornengo2} gravity effects
on neutrino mixing and oscillations have been discussed in connection with possible applications to
gravitational lensing.
On the other hand, in~\cite{ourextended} neutrino oscillations in the plane-wave formalism have been addressed in the framework of extended theories of gravity, showing that the neutrino phase can be sensitive to the violation of the strong equivalence principle (SEP). Such a principle states that any physical experiment (including gravitational physics) is locally not affected by the presence of a gravitational background field. This means that, if self-gravitational effects of the test bodies are not negligible but, at the same time, do not significantly influence the background gravitational field, then it is always possible to find local inertial observers who will find the same experimental result as if such a background field were absent~\cite{DiCasola:2013iia}. 
It is worth mentioning that SEP may be stated as the union of the Einstein equivalence principle (EEP) \textit{and} the gravitational weak equivalence principle (GWEP)~\cite{DiCasola:2013iia}. In particular, in this work we are interested in the violation of the latter, i.e., in scenarios in which self-gravity effects alter the motion of test bodies in some background gravitational field. In what follows, we simply refer to it as SEP, since a violation of GWEP would also imply a violation of SEP.

Motivated by the study of SEP violation in the neutrino plane-wave analysis of~\cite{ourextended}, here we discuss the natural extension to the more realistic wave-packet picture. Initially introduced in the context of neutrino flavor transitions to keep track of the coherence throughout the propagation~\cite{giuntiwp,giuntiwp2,giuntiwp3,kopp}, wave-packets have proven to be an important tool to characterize and isolate distinctive properties of both neutrino dynamics~\cite{capolupowp,capolupowp2} and the underlying
geometric background, especially in the presence of gravity~\cite{chatelain,comment,ouruniverse,hammad,gplwp,gplwp2}. Starting from a generic metric tensor that explicitly admits a violation of SEP, we show that the neutrino wave-packet dynamics acquires a dependence on the parameter that controls the amount of the said violation. Besides the theoretical interest, this result is also expected to have experimental implications as for the possibility to constrain extended theories of gravity. We observe that this analysis fits in a well-established research line aiming at testing the validity of the SEP via neutrino (and, more generally, particle physics)
phenomenology~\cite{neutrinowep2,neutrinowep3,neutrinowep4,neutrinowep5,neutrinowep6,neutrinowep7,neutrinowep8,neutrinowep}.


The remainder of the paper is organized as follows: 
first, we introduce the notion of SEP and 
review the wave-packet approach in the context of neutrino physics. After that, 
we investigate the impact of the SEP violation on the decoherence process due to the neutrino propagation in curved spacetime. For this purpose, 
we employ the weak-gravity approximation, which allows for a consistent identification of 
the physical distances and energies as measured by locally inertial observers~\cite{comment}. 
Furthermore, we assume neutrinos to be relativistic, as supported by empirical evidences~\cite{kopp}. 
Finally, 
we discuss our results and future perspectives. Throughout the work, we use natural units $\hbar=c=1$ and the mostly positive metric convention $\eta_{\mu\nu}=\mathrm{diag}(-1,1,1,1)$. 

\section{Parametrization of SEP violation}

Let us consider a general class of extended models of gravity, whose generic line element in the weak-field limit and in the static, spherically-symmetric case can be parameterized as
\be\label{metric}
{\rm d}s^2=-\lf[1+2\phi(r)\ri]{\rm d}t^2+\lf[1-2\psi(r)\ri]\lf({\rm d}r^2+r^2{\rm d}\Omega^2\ri),
\ee
where ${\rm d}\Omega^2={\rm d}\theta^2+\sin^2\theta {\rm d}\varphi^2$ 
is the angular part of the
metric on the two-sphere, while $\phi(r)$ and $\psi(r)$ denote the gravitational potentials. Henceforth, the radial dependence will be implicitly understood when omitted. 

To quantify violations of the SEP, we first introduce the Eddington-Robertson-Schiff parameter as~\cite{will,will2} 
\be
\label{ERS}
\gamma=\frac{\psi}{\phi}\,.
\ee 
In the following we will be concerned with the non-trivial case of  $\phi\neq\psi\Longrightarrow\gamma\neq1$.

Neglecting post-Newtonian contributions arising from exotic scenarios that contemplate the existence of anisotropies and/or preferred-frame effects, one can build the so-called Nordtvedt parameter $\eta$ as~\cite{nordtvedt,nordtvedt2,nordtvedt3}
\be\label{nord}
\eta=4\lf(\beta-1\ri)-\lf(\gamma-1\ri),
\ee
with $\beta$ containing the information on nonlinear gravitational effects~\cite{will}. The parameter $\eta$ is non-zero when the self-gravity of test bodies can affect their motion in a given background field and, therefore, it quantifies the violation of SEP according to our previous definition. 

As argued in Ref.~\cite{ourextended}, we reasonably assume that, in the weak-field limit, the nonlinearity is essentially ascribed to the leading-order terms which are represented by the general relativistic solution. Therefore, since $\beta_{\mathrm{GR}}=1$, we have 
\be\label{nord2}
\eta=1-\gamma=\frac{\phi-\psi}{\phi}\,.
\ee
The above relation provides the crucial ingredient of our next analysis. In particular, we will study
how the Nordtvedt parameter $\eta$ affects the decoherence of neutrino wave-packets by directly entering the definition of the density matrix associated with the particle state.


\section{Neutrino wave-packet treatment} 

According to the standard Pontecorvo's prescription, neutrino flavor states $|\nu_\alpha\ran$, $\alpha=e,\mu,\tau$, are given by 
coherent superpositions of the states with definite masses $|\nu_j\ran$, $j=1,2,3$~\cite{pontecorvo}. Assuming that a flavor neutrino propagates from a source $P$ to a detector $D$ in vacuum, we can describe the corresponding evolution in the wave-packet picture as
follows:
\begin{eqnarray}
\label{fstate}
|\nu_\al(x)\rangle&=&\sum_j U^*_{\al j}\psi_j(x)|\nu_j\rangle\,, \\[2mm] \psi_j(x)&=&\chi_j(\textbf{x})e^{-i\Phi_j(P,D)}  \, ,
\label{fstate2}
\end{eqnarray}
where $U_{\al j}$ denotes the generic element of the Pontecorvo-Maki-Nakagawa-Sakata mixing matrix~\cite{pontecorvo}.

Equation~\eqref{fstate} generalizes
the plane-wave treatment of~\cite{ourextended}. 
Following~\cite{kopp,chatelain}, we work with Gaussian-type wave-packets,
where the $j$-th wave-packet is set to be peaked around a given value $\textbf{p}_j$ in momentum space according to
\begin{eqnarray}
\label{fourier1}
\chi_j(\textbf{x})&=&\int\!\frac{{\rm d}^3p}{(2\pi)^3}e^{i\textbf{p}\cdot\textbf{x}}\chi_j(\textbf{p})\,,\\[2mm]
\label{fourier2}
\chi_j(\textbf{p})&=&\lf(\frac{2\pi}{\si_p^2}\ri)^{\frac{3}{4}}\!e^{-\frac{(\textbf{p}-\textbf{p}_j)^2}{4\si_p^2}}\,.
\end{eqnarray}
Here, $\si_p$ is the momentum-width of the Gaussian function, normalized in such a way that
\be\label{norm}
\int\frac{{\rm d}^3p}{(2\pi)^3}\lf|\chi_j(\textbf{p})\ri|^2=1\,.
\ee
In a generic curved spacetime, the quantum phase appearing in Eq.~\eqref{fstate} can be written as~\cite{stodolsky}
\be\label{phase}
\Phi_j(P,D)=\int_P^{D}p_\mu^{(j)}{\rm d}x^\mu\,,
\ee
where the wave-packet average momentum $p_\mu^{(j)}$ is defined along the path 
\be
\label{fm}
p_\mu^{(j)}=m_jg_{\mu\nu}\frac{{\rm d}x^\nu}{{\rm d}s}\,,
\ee
and $m_j$ is the $j$-th neutrino mass obeying $p_\mu^{(j)}p^{\mu}{}^{(j)}=(p^{(j)})^2=-m_j^2$.

For later convenience, we remark that the neutrino state $|\nu_\al(x)\rangle$ can be equivalently described in terms of the associated density matrix $\rho(x)$, which for pure states (just like the one considered above) is defined as~\cite{kopp}
\begin{eqnarray}\label{rho}
\rho(x)\!\!&=&\!\!|\nu_\al(x)\rangle\langle\nu_\al(x)|\\[2mm]
\nonumber
\!\!&=&\!\!\sum_{j,k}U^*_{\al j}U_{\al k}\psi_j(x)\psi^*_k(x)|\nu_j\rangle\langle\nu_k|
=\sum_{j,k}\rho_{jk}|\nu_j\rangle\langle\nu_k|\,,
\end{eqnarray}
where we have introduced the density matrix components $\rho_{jk}=U^*_{\al j}U_{\al k}\psi_j(x)\psi^*_k(x)$.

As we shall see in the next Section, the knowledge of the density matrix formalism turns out to be essential for the assessment of the wave-packet decoherence.   


\section{Decoherence and the strong equivalence principle} 

Without harming the generality of our next considerations, we follow~\cite{chatelain,comment,ouruniverse,gplwp,gplwp2} and focus 
on the case of neutrino radial propagation. In other terms, we set the angular coordinates to $\theta=\mathrm{const.}=\pi/2$ and $\varphi=\mathrm{const}$. 
Under these assumptions, one can show that the only non-trivial components of the four-momentum~\eqref{fm} with the metric~\eqref{metric} are
\begin{eqnarray}
\label{4m2}
p_t^{(j)}&=&-m_j(1+2\phi)\frac{{\rm d}t}{{\rm d}s}\,,  \\[2mm]
p_r^{(j)}&=&m_j(1-2\psi)\frac{{\rm d}r}{{\rm d}s}\,,
\label{4m3}
\end{eqnarray}
where $p_t^{(j)}=-E_j(\textbf{p})$ is the energy of the $j$-th mass eigenstate. Notice that this 
is a constant of motion, since the metric tensor~\eqref{metric} does not depend on $t$. 

Now, by means of the on-shell mass relation and the weak-field approximation, one can readily derive 
%
\be\label{step2}
\frac{{\rm d}r}{{\rm d}s}=\lf(1+\psi-\phi\ri)\sqrt{\frac{E_j^2(\textbf{p})}{m_j^2}-(1+2\phi)} \, ,
\ee
which can be employed to compute the expression of the covariant phase~\eqref{phase}. 
To this aim, we use the approximation of relativistic neutrinos, i.e., 
$m_j/E_j(\textbf{p})\ll1$~\cite{kopp}. In such a regime, we find
\begin{eqnarray}
&&\!\!\!\!\!\!\!\!\!\!\!\!\!\Phi_j(P,D)=-\int \hspace{-0.5mm}E_j(\textbf{p}){\rm d}t\nonumber
\\[2mm]
&&\!\!\!\!\!\!+\hspace{-0.5mm}\int(1-\phi-\psi)\hspace{-0.5mm}\lf[E_j(\textbf{p})-(1+2\phi)\frac{m_j^2}{2E_j(\textbf{p})}\ri]\hspace{-0.5mm}{\rm d}r. 
\label{step4}
\end{eqnarray}
By performing the above integrals, we are left with
\begin{align}\nonumber
\Phi_j(P,D)=&-E_j(\textbf{p})\lf(t_{DP}-d_{DP}\ri)\\[2mm] \label{covphase}
&-\frac{m_j^2}{2E_j(\textbf{p})}\lf(r_{DP}+r_{{SEP}}\ri)\,,
\end{align}
where we have used the shorthand notation
\begin{align}\nonumber
&t_{DP}=t_{D}-t_P\,, \quad d_{DP}=r_{DP}-\int_{r_P}^{r_{D}}\hspace{-1mm}\lf(\phi+\psi\ri){\rm d}r\,,\\[2mm] \label{notation}
&r_{DP}=r_{D}-r_P\,, \quad r_{{SEP}}=\int_{r_P}^{r_{D}}\phi\,\eta\,{\rm d}r\,.
\end{align}
It is worth stressing that the Nordtvedt parameter $\eta$ appears through $r_{SEP}$ in Eq.~\eqref{covphase} as an additional term that would vanish should the strong equivalence principle be satisfied.

Next, in compliance with~\cite{kopp,chatelain,comment}, we expand the energy $E_j(\textbf{p})$  around the peak $\textbf{p}_j$ of the Gaussian distribution in momentum space. This expansion allows us to neglect the intrinsic temporal spreading of the wave-packet throughout the propagation of the mixed particle. Note that a similar effect has no influence over the evolution of the neutrino coherence~\cite{kopp,chatelain,comment}. Hence, up to the leading order, one has 
\be
E_j(\textbf{p})\simeq E_j+(\textbf{p}-\textbf{p}_j)\cdot\textbf{v}_j\,,
\ee 
where $E_j$ is the $j$-th component of the energy evaluated at $\textbf{p}_j$, while $\textbf{v}_j$ is the group velocity of the $j$-th mass-eigenstate wave-packet. This latter quantity can be further manipulated by recalling that we are working in the relativistic limit, which yields the following expression for its modulus~\cite{chatelain}
\be\label{velocity}
v_j= |\textbf{v}_j|
=|\nabla_\textbf{p} E_j(\textbf{p})|_{\textbf{p}=\textbf{p}_j}
\simeq1-\frac{m_j^2}{2E_j^2} \, .
\ee
%
In order to gain insight on the flavor transition probability and the decoherence rate, we now have to determine the analytic expression of the density matrix elements $\rho_{jk}$ appearing in~\eqref{rho}. 
The considerations carried out so far bring forward a significant streamlining of the phase-shift $\Phi_{kj}=\Phi_k-\Phi_j$. By repeating for the $k$-th neutrino mass eigenstate the same considerations as for the $j$-th one, i.e., 
\begin{eqnarray}\label{qeq}
E_k(\textbf{q})&\simeq& E_k+(\textbf{q}-\textbf{p}_k)\cdot\textbf{v}_k\,,\nonumber\\[2mm]
v_k&=&
|\nabla_\textbf{q} E_k(\textbf{q})|_{\textbf{q}=\textbf{p}_k}
\simeq1-\frac{m_k^2}{2E_k^2} \,,
\end{eqnarray}
%
we can write
\begin{align}\nonumber
\!\!\Phi_{kj}&\!=E_{jk}(t_{DP}-d_{DP})+\lf(\frac{m_j^2}{2E_j}-\frac{m_k^2}{2E_k}\ri)\lf(r_{DP}+r_{{SEP}}\ri)\\[2mm] \label{phikj}
&\!\!\!+\textbf{v}_j\cdot(\textbf{p}-\textbf{p}_j)(t_{DP}-\ell_j)-\textbf{v}_k\cdot(\textbf{q}-\textbf{p}_k)(t_{DP}-\ell_k) \, ,
\end{align}
where
$E_{jk}=E_j-E_k$ and 
\be
\ell_{j}=\frac{m_{j}^2}{2E_{j}^2}\left(r_{DP}+r_{{SEP}}\right)+d_{DP}\,,
\ee
An analogous relation  holds for $\ell_{k}.$

By use of Eqs.~\eqref{fstate}, \eqref{fourier1} and \eqref{rho}, 
we now get  
\be\label{rhojk}
\rho_{jk}(x)=\Xi^\al_{jk}\int\frac{{\rm d}^3p}{(2\pi)^3}\int\frac{{\rm d}^3q}{(2\pi)^3}\,e^{i\Phi_{kj}}e^{-\frac{(\textbf{p}-\textbf{p}_j)^2}{4\si_p^2}}e^{-\frac{(\textbf{q}-\textbf{p}_k)^2}{4\si_p^2}} \,,
\ee
where we have introduced the shorthand notation 
\be
\label{shn}
\Xi^\al_{jk}=(2\pi/\si_p)^{3/2}U^*_{\al j}U_{\al k}\,.
\ee 
Integration of Eq.~\eqref{rhojk} over momenta gives
\begin{align}\nonumber
\rho_{jk}(x)&=\frac{\Xi^\al_{jk}}{(2\si_x\sqrt{\pi})^6}\exp\Bigl\{{iE_{jk}(t_{DP}-d_{DP})}\\[2mm] \nonumber
&+i\lf(\frac{m_j^2}{2E_j}-\frac{m_k^2}{2E_k}\ri)\lf(r_{DP}+r_{{SEP}}\ri)\\[2mm] \label{rhojk2}
&-{\si_p^2\bigl[\textbf{v}_k^2(t_{DP}-\ell_k)^2+\textbf{v}_j^2(t_{DP}-\ell_j)^2\bigr]}\Bigr\} \, ,
\end{align}
with $\si_x=1/2\si_p$ being the wave packet spatial width. One can check that the 
standard plane-wave picture is recovered
in the limit $\sigma_x\rightarrow\infty$, as expected. 

In neutrino oscillation experiments, the function that modulates the decoherent dynamics only depends on the spatial distance covered by the mixed particle~\cite{kopp}. Thus, to consistently account for this aspect, the quantity that should actually be considered is 
\be\label{fin}
\rho_{jk}(\textbf{x})=\int {\rm d}t\,\rho_{jk}(x) \, .
\ee
After a final integration~\cite{chatelain}, we note that $\rho_{jk}(\textbf{x})$ can be split in the product of three distinct contributions
\be\label{fin2}
\rho_{jk}(\textbf{x})=\xi^\al_{jk}\,\rho_{jk}^{\mathrm{osc}}(\textbf{x})\,\rho_{jk}^{\mathrm{damp}}(\textbf{x}) \,,
\ee
where the first term 
\be\label{kappa}
\xi^\al_{jk}=\frac{\sqrt{2}U^*_{\al j}U_{\al k}}{2\pi\si_x\bar{v}}\exp\lf[-\frac{E_{jk}^2\si_x^2}{\bar{v}^2}\ri],\quad\, \bar{v}=\sqrt{\textbf{v}_j^2+\textbf{v}_k^2}\,,
\ee
does not influence the flavor transition probability and decoherence rate. On the other hand, the second factor reads
\begin{align}\nonumber
\rho_{jk}^{\mathrm{osc}}=&\exp\lf\{i \, \lf(r_{DP}+r_{{SEP}}\ri)\, \lf[\lf(\frac{m_j^2}{2E_j}-\frac{m_k^2}{2E_k}\ri)\ri.\ri.\\[2mm] \label{rosc}
&\lf.\lf.-\frac{E_{jk}}{\bar{v}^2}\lf(\textbf{v}_j^2\frac{m_j^2}{2E_j^2}+\textbf{v}_k^2\frac{m_k^2}{2E_k^2}\ri)\ri]\ri\} \, ,
\end{align}
which, as shown in~\cite{ourextended}, can be used to compute the  flavor transition probability $j\leftrightarrow k$.

For the main purpose of the present work,  the relevant contribution comes from the third factor, namely $\rho_{jk}^{\mathrm{damp}}(\textbf{x})$, which carries the information on the decoherence mechanism experienced by the flavor neutrino along its propagation. By writing the said quantity up to the leading-order terms in $m_j/E_j$ and $m_k/E_k$, and working in the typical approximation $E_j\simeq E_k\equiv E,$ one obtains
\bea
\rho_{jk}^{\mathrm{damp}}(\textbf{x}) & = & \exp\lf[-\frac{\Delta m_{kj}^4\lf(r_{DP}+r_{{SEP}}\ri)^2}{32\,\si_x^2\,E^4}\ri] \non \\[2mm] \label{rdamp}
& = & \rho_{jk}^{\mathrm{damp},0}(\textbf{x})\rho_{jk}^{\mathrm{damp},\eta}(\textbf{x}) \, , 
\eea
where $\Delta m_{kj}^2=m_k^2-m_j^2$, while $E$ can be interpreted as the average transition energy between the $j$-th and the $k$-th mass eigenstate.
For a clearer presentation of the result, in the second line of~\eqref{rdamp} we have factorized $\rho^{\mathrm{damp}}$ as the product of 
\be
\rho_{jk}^{\mathrm{damp},0}(\textbf{x}) \ = \ \exp\lf[-\frac{\Delta m_{kj}^4 r_{DP}^2}{32\,\si_x^2\,E^4}\ri] \, , 
\ee
which is the standard factor arising in Einstein's general relativity and
\be
\rho_{jk}^{\mathrm{damp},\eta}(\textbf{x}) \ = \ \exp\lf[-\frac{\Delta m_{kj}^4\, r_{SEP} \lf(r_{SEP}+2 r_{DP}\ri)}{32\,\si_x^2\,E^4}\ri] \, , 
\ee
which is the correction we gain in gravitational theories where the strong equivalence principle is violated. Clearly, for $\eta\to 0$ we have $r_{SEP}\to 0$ from Eq.~\eqref{notation}, which in turn implies $\rho_{jk}^{\mathrm{damp},\eta}\to 1$, as expected. In this case, one can also introduce the standard definition of neutrino coherence length $L_{coh}$ as the distance at which the density matrix is suppressed by a factor $e^{-1}$, i.e.~\cite{giuntiwp2} 
\be
\label{lcoh}
L^{k j}_{coh}=\frac{4\sqrt{2}E^2\,\sigma_x}{\Delta m_{kj}^2}\,,
\ee
while the \emph{oscillation length} is defined as usual by~\cite{giuntibook}
\be \label{losc}
L^{k j}_{osc} \ = \ \frac{4\pi E}{\Delta m_{kj}^2} \, .
\ee
On a final note, considering the diagonal elements of the density matrix, one can compute the (normalized) survival probability, which for an electron neutrino 
turns out to be
\begin{eqnarray}\nonumber
\!P_{\nu_e \rightarrow \nu_e}(\G x)&\hspace{-1mm}=\hspace{-1mm}&\sum_{j,k} |U_{e j}|^2|U_{e k}|^2\exp\Bigl[i\lf(r_{DP}+r_{{SEP}}\ri)\frac{\De m_{kj}^2}{2E}\\[2mm]
&&\!\!\!\!\hspace{20mm}-\,\frac{\Delta m_{kj}^4\lf(r_{DP}+r_{{SEP}}\ri)^2}{32\,\si_x^2\,E^4}\Bigr].
\end{eqnarray}
{By bringing out an overall $r_{DP}$  and recognizing that $r_{SEP}/r_{DP}$ is nothing but the average value of $\phi\hspace{0.4mm} \eta$,
we recast the above equation in the form
\begin{eqnarray}\nonumber
\!P_{\nu_e \rightarrow \nu_e}(\G x)&\hspace{-1mm}=\hspace{-1mm}&\sum_{j,k} |U_{e j}|^2|U_{e k}|^2\exp\Bigl[i r_{DP} \xi\frac{\De m_{kj}^2}{2E}\\[2mm]
&&\!\!\!\!\hspace{20mm}-\,\frac{\Delta m_{kj}^4 r_{DP}^2 \xi^2}{32\,\si_x^2\,E^4}\Bigr] \, , 
\label{ProbSmald}
\end{eqnarray}
with $\xi \equiv\lf(1+\frac{r_{SEP}}{r_{DP}}\ri)$.
From this relation, we infer that the modified coherence and oscillation lengths read
\begin{eqnarray}
\label{modcohl}
L^{kj,SEP}_{coh}&=&\frac{4\sqrt{2}E^2\,\sigma_x}{\Delta m_{kj}^2\,\xi}\simeq\frac{4\sqrt{2}E^2\,\sigma_x}{\Delta m_{kj}^2}\left(1-\frac{r_{SEP}}{r_{DP}}\right)\,,\,\,\,\,\,\,\,\,\,\,\\[2mm]
L^{kj,SEP}_{osc}&=&\frac{4\pi E}{\Delta m_{kj}^2\,\xi}\simeq\frac{4\pi E}{\Delta m_{kj}^2}\left(1-\frac{r_{SEP}}{r_{DP}}\right)\,,
\label{loscsep}
\end{eqnarray}
where, by exploiting the weak field approximation, in the second step we have expanded up to the leading-order term in $r_{SEP}/r_{DP}$, which is expected to be small. We shall verify this assumption \emph{a posteriori}.  
}

{A comment is in order here. If we set $r_{SEP}=0$ in the above formulas, e.g. in Eq.~\eqref{ProbSmald}, the dependence on the gravitational potentials $\Phi$ and $\Psi$ seem to disappear completely, thus apparently entailing that the gravitational implications are non-vanishing only when $r_{SEP}\neq 0.$ However, this reasoning does not account for the fact that the dependence on the gravitational coupling is implicit in the definition of the coordinate distance $r_{DP}$ and $E$, which is the energy measured by an observer at infinity.} {Indeed, to render the gravitational contribution explicit, we have to
introduce the energy $E_\ell$ measured by a local observer at rest with the experiment and the proper length $L_p$ traveled by the neutrino, {which are the true physical quantities detected in a local experiment.} These are related to the asymptotic energy $E$ and the coordinate distance $r_{DP}$ by the following relations:
\bea
\label{Newp}
E_\ell\  = \ [1-\phi(r)] \, E \, , \\[2mm]
L_p \ = \ r_{DP}-\int^{r_D}_{r_P} \!\! \dr r \, \psi(r) \,.
\eea
Hence, as shown in a significant number of works (see for instance~\cite{fornengo,fornengo2,Godunov:2009ce,ourextended}), the gravitational corrections can be made manifest by rephrasing the relevant observable quantities in terms of $E_\ell$ and $L_{p}.$} 

{For instance, by virtue of the aforementioned substitutions, the probability~\eqref{ProbSmald} can be written as
\begin{eqnarray}
&&\!\!\!\!\!\!\!\!\!\!\!P_{\nu_e \rightarrow \nu_e}(\G x)=\sum_{j,k} |U_{e j}|^2|U_{e k}|^2\times\nonumber\\[2mm]
&&\!\!\!\!\!\times \exp\left[i \frac{\De m_{kj}^2 L_p\xi}{2E_\ell}\left( 1-\phi(r_D)+\frac{1}{L_p}\int_{r_P}^{r_D}{\rm d}r\, \psi(r) \right)\right. \nonumber\\[2mm]
&&\!\!\!\! \!\left.- \frac{\Delta m_{kj}^4 L_p^2 \xi^2}{32\,\si_x^2\,E_\ell^4}\left( 1-4\phi(r_D) +\frac{2}{L_p}\int_{r_P}^{r_D}{\rm d}r\,\psi(r)\right) \right].
\label{ProbSmald-local-quant}
\end{eqnarray}
In the case of Einstein's general relativity, i.e. $\phi(r)=\psi(r)=-Gm/r$ ($\eta=0,$ $r_{SEP}=0$ and $\xi=1$), the formula~\eqref{ProbSmald-local-quant} reads
\begin{eqnarray}
&&\!\!\!\!\!\!\!\!\!\!\!P_{\nu_e \rightarrow \nu_e}(\G x)=\sum_{j,k} |U_{e j}|^2|U_{e k}|^2\times\nonumber\\[2mm]
&&\!\!\!\!\!\times \exp\left[i \frac{\De m_{kj}^2 L_p}{2E_\ell}\left( 1+\frac{Gm}{r_D}-\frac{Gm}{L_p}\log\left(\frac{r_D}{r_P}\right) \right)\right. \nonumber\\[2mm]
&&\!\!\!\! \!\left.- \frac{\Delta m_{kj}^4 L_p^2}{32\,\si_x^2\,E_\ell^4}\left( 1+\frac{4Gm}{r_D} -\frac{2}{L_p}\log\left(\frac{r_D}{r_P}\right)\right) \right]\,,
\label{ProbSmald-local-quant-GR}
\end{eqnarray}
where it is evident that the gravitational contribution is non-vanishing even when $r_{SEP}=0.$ When gravity is switched off, i.e., setting $G\to0,$ we recover the Minkowski scenario, as expected. Moreover, one can easily verify that, in the limit $\sigma_x\rightarrow \infty$, the well-known result of the plane-wave picture~\cite{fornengo,fornengo2,Godunov:2009ce,ourextended} is consistently recovered.
}

{Before concluding this section, let us emphasize that the expressions~\eqref{ProbSmald} and~\eqref{ProbSmald-local-quant} for the survival probability} convey that, from an experimental perspective, the presence of $\eta$ modifies the expected flux of neutrinos with respect to the general relativistic prediction \cite{Capozziello:1999qm}. This can be easily understood by observing that the survival probability for a fixed flavor, say $\nu_\sigma$, can be estimated as $P_{\nu_\sigma\rightarrow\nu_\sigma}={\Phi_{\nu_\sigma}}/{\Phi_{tot}}$, where $\Phi_{tot}=\sum_{\alpha}\Phi_{\nu_{\alpha}}$ is the total flux of neutrinos from a given source, while $\Phi_{\nu_{\sigma}}$ is the measured flux of neutrinos $\nu_\sigma$~\cite{Berez}. 

\subsection{Experimental constraint}\label{subsec-exp}

{For practical purposes, it is interesting to constrain the correction $r_{SEP}/r_{DP}$ by addressing a realistic experimental setup. As an example, we consider the case of solar neutrinos produced by the pp solar fusion cycle ($E\lesssim 0.4\,\mathrm{MeV}$ and uncertainty $\sim 2\hspace{-0.3mm}-\hspace{-0.3mm}3\%$, $\Delta m_{sol}^2\equiv\Delta m_{12}^2=(0.75\,\pm\,0.02)\times10^{-4}\,\mathrm{eV}^2$)~\cite{PDG}. Since at present we do not have direct measurements of the neutrino wave-packet spread $\sigma_x$ but only upper bounds from reactor neutrino experiments~\cite{JUNO:2021ydg,DeRomeri:2023dht}, a preliminary result consists in constraining the SEP parameter by looking at its correction to the effective oscillation length~\eqref{loscsep}. By taking into account that SEP violation has not been spotted in experiments, we impose that the correction in Eq.~\eqref{loscsep} lies within the margin of the error on measurements of $L_{osc}$, i.e., 
\be
\label{bound}
\frac{4\pi E}{\Delta m_{sol}^2}\, \frac{r_{SEP}}{r_{DP}}\lesssim \mathcal{O}(10^9)\,\mathrm{eV}^{-1}\,,
\ee
}
{where the error on the oscillation length has been estimated from Eq.~\eqref{losc} using error propagation. In light of this, we get $r_{SEP}/r_{DP}\lesssim\mathcal{O}(10^{-6})$,
which substantiates the validity of our leading-order expansion in Eqs.~\eqref{modcohl} and~\eqref{loscsep}. While being relatively small, this result does not rule out the possibility of future detection of SEP violation in the next-generation neutrino experiments~\cite{Gann}. 
}

\subsection{Discussion} 

{From the expressions of the density matrix and survival probability derived above, we can also note that}, when $\eta\neq 0,$ a violation of the SEP affects the damping of the neutrino wave-packets in a way that  depends on the sign of $r_{SEP}$: if $r_{SEP}>0$ the damping is stronger than the standard general relativity ($r_{SEP}=0$) description, while in the opposite case it is weaker (see Eq.~\eqref{rdamp}). 

To discuss our outcome more concretely, we can consider an explicit example in which $\eta$ can be non-zero. Specifically, we focus on metric potentials corrected by Yukawa-type contributions of the form
\begin{eqnarray}
\label{f(R)-pot1}
\phi&=&-\frac{GM}{r}\left(1+a\,e^{-r/\lambda}\right)\,,\\[2mm] 
\psi&=&-\frac{GM}{r}\left(1+b\,e^{-r/\lambda}\right) \, , \label{f(R)-pot}
\end{eqnarray}
where $G$ is Newton's constant, $M$ the mass of the source, such as a star or a planet, and $\lambda$  the length scale at which the effects of the exponential terms become important. The constants $a$ and $b$ 
are model-dependent.

Given the potentials~\eqref{f(R)-pot1} and~\eqref{f(R)-pot}, from the definition~\eqref{notation} we obtain
\be
r_{SEP}=(a-b)GM[-{\rm Ei}(-r_D/\lambda)+{\rm Ei}(-r_P/\lambda)] \,, \label{rsep-f(R)}
\ee
where the overall sign is determined by the signs of $a-b$ and $r_D-r_P;$ the special function ${\rm Ei}(-x)=-\int_{x}^{\infty}{\rm d}t\,e^{-t}/t$ is the so-called exponential integral. If we assume that the neutrinos are emitted by the gravitational source, then $r_D>r_P$ and the term in the square brackets in Eq.~\eqref{rsep-f(R)} is always negative, since ${\rm Ei}(-x)$ is a monotonically increasing function of $x>0.$ Therefore, in this case we get $r_{SEP}>0$ ($r_{SEP}<0$) for $a<b$ ($a>b$).

Yukawa-like corrections are very common in scenarios admitting the existence of a fifth force. A popular theory in which the two linear potentials are fully known is $f(R)$ gravity~\cite{DeFelice:2010aj}, in which the weak-field limit gives $a=-b=1/3,$ which means that $a>b.$ For this model, the role of the fifth force is played by an additional massive scalar degree of freedom of mass $m=1/\lambda.$ Therefore, in this theory the violation of the strong equivalence principle manifests itself as a weaker damping, i.e., a less damped neutrino wave-packet, as compared to the case of Einstein's general relativity. 

The above scenario includes only a specific example; however, the same steps can be easily applied to any other gravitational theory~\cite{capoz1} for which the linear metric potentials are known.

{On a final note, one could tailor the phenomenological considerations in Sec.~\ref{subsec-exp} to the above model and try to constrain the length scale $\lambda$ of Yukawa-type deviations from Newtonian gravity. However, with reference to the solar neutrino setup described above, the ensuing bound would not improve the best experimental limit $\lambda\sim\mathcal O(10)\,\mathrm{\mu m}$ obtained via torsion-balance~\cite{TB} and fiber interferometer~\cite{fiber} tests. This was somehow predictable in the weak-field approximation we are considering here. Furthermore, the same holds for any reverse attempt to estimate $r_{SEP}/r_{DP}$ from known upper bounds on $\lambda$. 
On the other hand, promising results are expected when working in the strong-gravity regime, e.g. neutrinos emitted by supernova explosions or scattered by black holes. This analysis represents a natural upgrade of the present study and will be further developed elsewhere.}

\section{Concluding remarks}
Motivated by the analysis of~\cite{ourextended}, we have shown that violations of the strong equivalence principle can be witnessed in neutrino oscillations both through the flavor oscillation phase and the damping of neutrino wave-packets; indeed, such quantities explicitly depend on the Nordtvedt parameter $\eta$. This result has been achieved in the case of a weak-field, spherically symmetric metric tensor and can be applied to generic extended theories of gravity for which the  potentials are known.

Let us stress that, although here we discussed the case of neutrino oscillations in vacuum, the same treatment can be easily extended to the case of oscillations in matter~\cite{Wolfenstein1978,Smirnov1985} by
a suitable (background-dependent) redefinition of the neutrino mass-shift and 
metric potential $\phi$.
In fact, as it was shown for plane-waves~\cite{ourextended}, the presence of matter does not seem to qualitatively change the final outcomes.

Apart from the theoretical interest, it has been remarked that nontrivial gravitational contributions could significantly affect neutrino oscillation in supernova explosions~\cite{PhysRevD.55.2760} and solar neutrino experiments~\cite{Capozziello:1999qm}, thus giving rise to effects that are potentially measaurable in the next-generation neutrino esperiments. 

\vspace{3mm}

\section*{Acknowledgements}
Nordita is supported in part by NordForsk.
L.P. acknowledges support by MUR (Ministero dell'Universit\`a e della Ricerca) via the project PRIN 2017 ``Taming complexity via QUantum Strategies: a Hybrid Integrated Photonic approach'' (QUSHIP) Id. 2017SRNBRK and is thankful to the ``Angelo Della Riccia'' foundation for the awarded fellowship received to support the study at Universit\"at Ulm. L.S. was supported by the Polish National Science Center grant 2018/31/D/ST2/02048. G.G.L. acknowledges the Spanish ``Ministerio de Universidades''
for the awarded Maria Zambrano fellowship and funding received
from the European Union - NextGenerationEU.
L.S and G.G.L. are also grateful for participation to the  LISA Cosmology Working group.
G.G.L. and L.P. acknowledge  networking support by the COST Action CA18108.


\bibliography{LibraryNeutrinograv}

\bibliographystyle{apsrev4-2}

\end{document}